\documentclass[aps,prl,twocolumn,byrevtex,twocoloumn,byrevtex,floatfix,showpacs,superscriptaddress]{revtex4}

\usepackage{amssymb,graphicx}
\usepackage{color}
\begin{document}

\title{Unidirectional terahertz light absorption in the pyroelectric ferrimagnet CaBaCo$_4$O$_7$}

\author{S. Bord\'acs}
\affiliation{Department of Physics, Budapest University of Technology and Economics and MTA-BME Lend\"ulet Magneto-optical Spectroscopy Research Group, 1111 Budapest, Hungary}
\affiliation{Department of Applied Physics and Quantum Phase Electronics Center (QPEC), University of Tokyo, Tokyo 113-8656, Japan}

\author{V. Kocsis}
\affiliation{Department of Physics, Budapest University of Technology and Economics and MTA-BME Lend\"ulet Magneto-optical Spectroscopy Research Group, 1111 Budapest, Hungary}

\author{Y. Tokunaga}
\affiliation{RIKEN Center for Emergent Matter Science (CEMS), Wako 351-0198, Japan}
\affiliation{Department of Advanced Materials Science, University of Tokyo, Kashiwa 277-8561, Japan}

\author{U. Nagel}
\affiliation{National Institute of Chemical Physics and Biophysics, 12618 Tallinn, Estonia}

\author{T. R\~o\~om}
\affiliation{National Institute of Chemical Physics and Biophysics, 12618 Tallinn, Estonia}

\author{Y. Takahashi}
\affiliation{Department of Applied Physics and Quantum Phase Electronics Center (QPEC), University of Tokyo, Tokyo 113-8656, Japan}
\affiliation{RIKEN Center for Emergent Matter Science (CEMS), Wako 351-0198, Japan}

\author{Y. Taguchi}
\affiliation{RIKEN Center for Emergent Matter Science (CEMS), Wako 351-0198, Japan}

\author{Y. Tokura}
\affiliation{Department of Applied Physics and Quantum Phase Electronics Center (QPEC), University of Tokyo, Tokyo 113-8656, Japan}
\affiliation{RIKEN Center for Emergent Matter Science (CEMS), Wako 351-0198, Japan}

\date{\today}

\begin{abstract}
Spin excitations were studied by absorption spectroscopy in CaBaCo$_4$O$_7$ which is a type-I multiferroic compound with the largest magnetic-order induced ferroelectric polarization ($\Delta$$P$=17\,mC/m$^2$) reported, so far. We observed two optical magnon branches: a solely electric dipole allowed one and a mixed magnetoelectric resonance. The entangled magnetization and polarization dynamics of the magnetoelectric resonance gives rise to unidirectional light absorption, i.e.~that magnon mode absorbs the electromagnetic radiation for one propagation direction but not for the opposite direction. Our systematic study of the magnetic field and temperature dependence of magnon modes provides information about the energies and symmetries of spin excitations, which is required to develop a microscopic spin model of CaBaCo$_4$O$_7$.
\end{abstract}

\maketitle

Multiferroic materials have gained tremendous interest because of their versatile possibilities for applications in sensing, data storage and computing~\cite{Gajek2007,Matsukura2015,Wang2009}. These functionalities rely on the magneto-electric (ME) effect which is enhanced in multiferroics by the coexistence of ferroelectric and magnetic order. Recent studies of multiferroic materials from gigahertz frequencies to X-ray wavelengths have demonstrated that finite frequency ME effect gives rise to another useful phenomenon, the directional dichroism~\cite{Jung2004,Kubota2004,Arima2008Ome,Kezsmarki2011,Bordacs2012,Takahashi2012,Takahashi2013,Kezsmarki2014,Kibayashi2014,Kuzmenko2014,KezsmarkiBFO2015}. The non-reciprocal directional dichroism and its Kramers-Kronig counterpart, non-reciprocal directional birefringence, are the absorption coefficient and refractive index differences, respectively, for counter-propagating light beams detectable irrespective of the light polarization state. These effects enable new applications for multiferroics in photonics, e.g.~one-way light guides and polarization rotators. The largest directional dichroism in multiferroics is at terahertz (THz) frequencies~\cite{Kezsmarki2014} what can boost terahertz photonics in the future, similar to the prospect of graphene non-reciprocal terahertz devices~\cite{Tamagnone2014}. The dc and the finite frequency ME effect, which determine the efficiency of any multiferroic device, are intimately connected by a sum rule~\cite{Szaller2014} allowing a large static ME effect only if a large non-reciprocal directional dichroism is present in the absorption spectrum of electromagnetic radiation. The pyroelectric ferrimagnet CaBaCo$_4$O$_7$ is a type-I multiferroic compound with the largest magnetic-order induced ferroelectric polarization ($\Delta$$P$=17\,mC/m$^2$) reported, so far~\cite{Caignaert2013}. In this paper we report a magnon mode in CaBaCo$_4$O$_7$ which shows unidirectional light absorption. Namely, the non-reciprocal directional dichroism is so strong for the magnon resonance that it absorbs terahertz radiation only in one direction, but not in the opposite direction.

In the long wavelength limit the directional dichroism emerges from the dynamic magnetoelectric (ME) coupling, $\delta${\bf P}$_\omega$=$\hat{\chi}^{em}$($\omega$){\bf H}$_\omega$ and $\delta${\bf M}$_\omega$=$\hat{\chi}^{me}${\bf E}$_\omega$~\cite{Kezsmarki2011,Bordacs2012}. The induced polarization, $\delta${\bf P}$_\omega$ and magnetization, $\delta${\bf M}$_\omega$ interfere constructively/destructively with {\bf H}$_\omega$ and {\bf E}$_\omega$ fields of the light as the relative orientation of {\bf H}$_\omega$ and {\bf E}$_\omega$ is opposite for foreward/backward propagating light beams~\cite{Barron2004}. Due to the strong hybridization between the electric and magnetic dipole excitations large directional dichroism has been found recently for the collective spin excitations of multiferroics. Non-reciprocal directional dichroism has been observed in two distinct cases, so far. The first one, the so-called \emph{magnetochiral dichroism} (MChD), requires a chiral material with finite magnetization, and the directional dichroism is present for light beams propagating parallel and antiparallel to the direction of the magnetization. MChD has been detected in several akermanite compounds, (Ba$_2$CoGe$_2$O$_7$, Sr$_2$CoSi$_2$O$_7$, Ca$_2$CoSi$_2$O$_7$)~\cite {Bordacs2012,Kezsmarki2014} and in Cu(Fe,Ga)O$_2$~\cite{Kibayashi2014}. The second one is realized when a material simultaneously possesses magnetization, {\bf M} and electric polarization (or a polar axis), {\bf P} \cite{note1}. In this case, the absorption is different for beams propagating along and opposite to the toroidal moment, which is the cross product of the electric polarization and the magnetization, {\bf T}={\bf P}$\times${\bf M}. We term the latter effect as \emph{toroidal dichroism} (TD), which has been observed in Ba$_2$CoGe$_2$O$_7$~\cite{Kezsmarki2011}, (Eu,Y)MnO$_3$~\cite{Takahashi2012}, (Gd,Tb)MnO$_3$~\cite{Takahashi2013}. All the above listed compounds are type-II multiferroics, where the usually weak ferroelectricity appears below the magnetic phase transition. On the other hand, type-I multiferroics usually have large ferroelectric polarization which is already present in their paramagnetic phase~\cite{Wang2009}. Moreover, they often have higher magnetic ordering temperature which enables room temperature applications. However, in this class of compounds, the dynamic magnetoelectric effect of spin-wave excitations has been studied only very recently in BiFeO$_3$~\cite{KezsmarkiBFO2015}, and information about the non-reciprocal directional dichroism is still scarce. Here, we focus on the non-reciprocal directional dichroism spectrum of a type-I multiferroic material, CaBaCo$_4$O$_7$, which belongs to the Swedenborgite family having members with multiferroic phases close to room-temperature~\cite{Raveau2008}.

\begin{figure}[h!]
\includegraphics[trim=0 3 3 4, clip, width=3.3in]{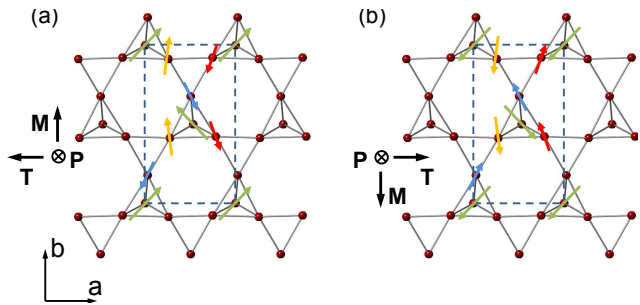}
\caption{(Color online) The magnetic order in CaBaCo$_4$O$_7$ which develops on the kagome and triangular lattice layers of Co ions~\cite{Caignaert2010}. The different colors indicate four different magnetic moments of inequivalent Co ions. The (a) and (b) panels show two magnetic domains with opposite magnetization, {\bf M} (up/down) and opposite toroidal moment, {\bf T} (left/right).}
\label{fig1}
\end{figure}

At room temperature CaBaCo$_4$O$_7$ has an orthorhombic, non-centrosymmetric crystal structure (space group $Pbn2_1$)~\cite{Caignaert2009}, which allows electric polarization along the $c$ axis. The tetrahedrally coordinated cobalt ions, responsible for magnetism of this compound, form alternating kagome and triangular lattice layers as shown in Fig.~\ref{fig1}. These sublattices with the strong antiferromagnetic exchange coupling between the cobalt ions are textbook examples of geometrical frustrated systems. In CaBaCo$_4$O$_7$ an orthorhombic distortion releases the frustration, and a ferrimagnetic order develops below $T$$_c$=60\,K (see Fig.~\ref{fig1})~\cite{Caignaert2009,Caignaert2010}. The $c$ axis is the hard magnetization axis, while the magnetization anisotropy within the $ab$ plane is small if any~\cite{Iwamoto2012}. The magnetic unit cell, which is equivalent to the structural unit cell, contains 16 cobalt spins, and the symmetry of the ferrimagnetic state reduces to $Pb'n2'_1$. Upon the magnetic phase transition a peak-like anomaly shows up in the temperature dependence of the dielectric function, indicating a coupling between fluctuating spin and charge degrees of freedom~\cite{Iwamoto2012,Caignaert2013}. Pyrocurrent measurements revealed a magnetic-order induced electric polarization, $\Delta$$P$$\approx$17\,mC/m$^2$ being the largest observed among multiferroics known, so far~\cite{Caignaert2013}. First principles calculations suggest that magnetostriction is the dominant driving mechanism of the large change in the ferroelectric polarization~\cite{Johnson2014}.

\begin{figure}[h!]
\vspace{6pt}
\includegraphics[width=3.3in]{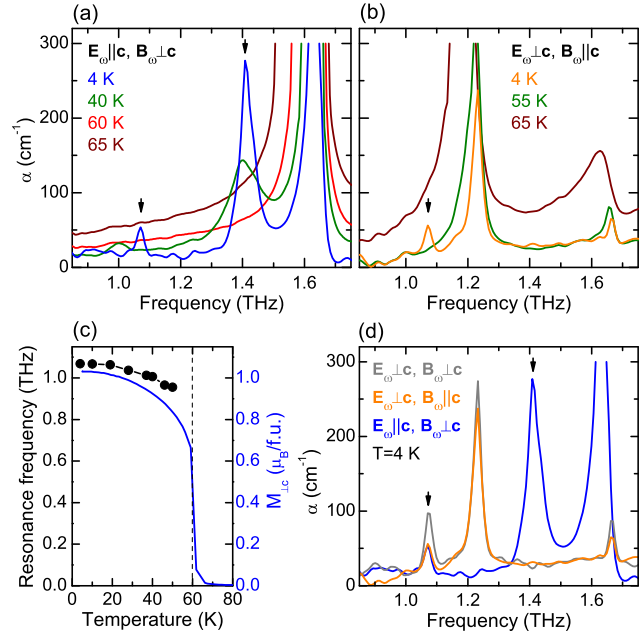}
\caption{(Color online) (a)-(b) The temperature dependence of the low-frequency absorption spectrum in CaBaCo$_4$O$_7$. The magnon modes at 1.07\,THz and 1.41\,THz, indicated by black arrows, disappear above the magnetic ordering temperature, $T$$_c$=60\,K. (c) The temperature dependence of the resonance frequency of the magnetoelectric resonance (filled circles) and the in-plane magnetization in $B$=0.1\,T magnetic field (solid line). (d) The polarization selection rules for the THz excitations of CaBaCo$_4$O$_7$. The 1.07\,THz excitation is sensitive to the orientation of both the electric and magnetic components of the light, while the magnetic mode at 1.41\,THz can be excited only by the electric field of the radiation when {\bf E}$_\omega$$\parallel$$\mathbf{c}$. The absorption spectra were cut when the absorption coefficient exceeded the detection limit.}
\label{fig2}
\end{figure}

%\section{Experimental}

Single crystals of CaBaCo$_4$O$_7$ were grown by the floating-zone technique~\cite{Caignaert2013}. The room temperature X-ray diffraction pattern was consistent with the space group $Pbn2_1$. X-ray diffraction and polarized optical microscopy imaging revealed that orthorhombic twinning occurs on microscopic scale in the studied samples. THz transmission spectra were measured on 300-500\,$\mu$m thick platelet-like samples with different crystallographic orientations using time-domain THz~\cite{Kida2004} and Fourier-transform infrared spectroscopy~\cite{Room2004}.

%\section{Results and discussion}

The THz spectra of CaBaCo$_4$O$_7$ measured with linear light polarization along the principal crystallographic axes using Fourier-transform spectroscopy are shown in Fig.~\ref{fig2}. Since the $a$ and $b$ axis are indistinguishable due to the orthorhombic twinning, the in-plane directions are labeled as $\perp$$c$. The temperature dependence of the THz spectra (see Fig.~\ref{fig2}(a)-(b)) shows that beside the temperature independent phonon modes observed in the paramagnetic phase two new resonances appear below $T$$_c$=60\,K whose resonance frequencies are shifted to lower frequencies as the temperature increased (Fig.~\ref{fig2}(c)). In the low temperature phase no additional infrared active phonon mode is expected since (i) the magnetic and the structural unit cells are the same and (ii) the space group changes from $Pbn2_1$ to $Pb'n2'_1$ upon the magnetic phase transition, while keeping all spatial symmetries intact. Therefore, the two resonances, located at 1.07\,THz and 1.41\,THz at $T$=4\,K can be identified as magnetic excitations. These peaks shift to lower frequencies and become broader as the temperature increases as expected for collective excitations of a magnetic order.

The selection rules for the low-frequency excitations were obtained from the light polarization dependent measurements, see Fig.~\ref{fig2}(d). The modes at 1.23\,THz and 1.67\,THz are $ab$-plane lattice vibrations, while the 1.63\,THz resonance is a $c$-axis optical phonon mode since all of these excitations are sensitive only for the orientation of {\bf E}$_\omega$ and they are also present in the paramagnetic phase above $T_c$ as shown in Fig.~\ref{fig2}(a)-(b). However, the intensity of the 1.07\,THz resonance changes when either $\mathbf{B}$$_\omega$ or $\mathbf{E}$$_\omega$ is rotated out from the $ab$ plane, therefore, this mode is assigned to a both electric and magnetic dipole active resonance excitation. The magnetic mode at 1.41\,THz can be excited only with {\bf E}$_\omega$$\parallel$$\mathbf{c}$, thus, it is an electric-dipole allowed but magnetic-dipole forbidden magnetic resonance. Such strong electric dipole active electromagnons have been found also in orthorhomic manganites~\cite{Pimenov2006,Takahashi2012}.

\begin{figure}[h!]
\vspace{10pt}
\includegraphics[width=2.5in]{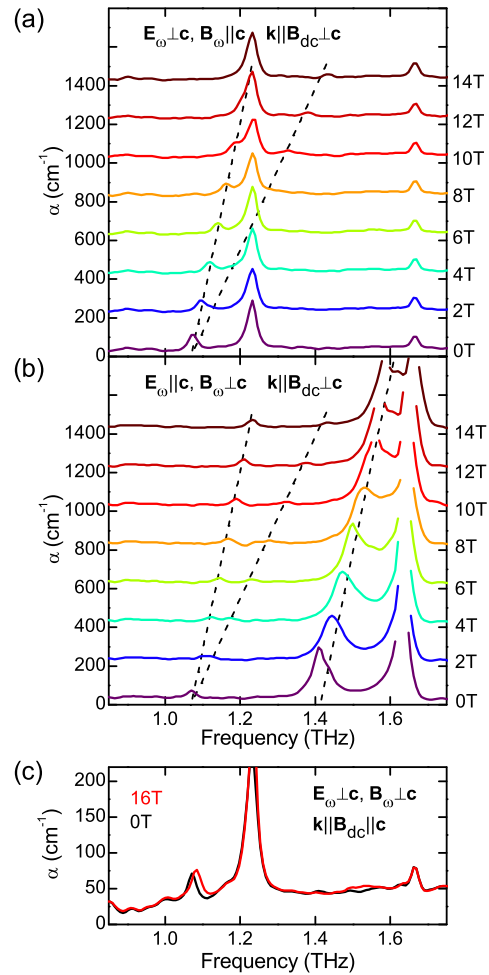}\\
\caption{(Color online) (a)-(c) Solid lines, the magnetic field dependence of the absorption spectra of CaBaCo$_4$O$_7$ at 4\,K. The spectra are shifted vertically proportional to the magnitude of the field in panels (a) and (b). Dashed lines, the evolution of the resonance frequencies in field.}
\label{fig3}
\end{figure}

The magnetic field dependence of the THz spectra was measured for different field orientations at 4\,K as shown in Fig.~\ref{fig3}. The experiments were performed in Faraday geometry, where the external magnetic field is parallel to the light propagation direction, {\bf k}$\parallel${\bf B}$_{dc}$, thus no directional dichroism is expected by symmetry. When the magnetic field lies within the $ab$ plane, {\bf B}$_{dc}$$\perp$$\mathbf{c}$ (see Fig.~\ref{fig3}(a)-(b)), the magnetoelectric mode at 1.07\,THz is asymmetrically split into two branches, which get separated linearly by 25\,GHz/T, while the 1.41\,THz resonance is shifted to higher frequencies by about 13\,GHz/T. This strong magnetic field dependence also supports the above assignment of 1.07\,THz and 1.41\,THz resonances to magnetic excitations. The other lines are infrared active phonon modes showing no magnetic field dependence. When an external magnetic field of 16\,T is applied parallel to the $c$ axis, the spectrum does not change significantly (Fig.~\ref{fig3}(c)); only the 1.07\,THz resonance is shifted slightly to higher frequencies and the 1.41\,THz mode becomes weakly active. This result suggests that the magnetic structure can be hardly modified by $\mathbf{B}$$_{dc}$$\parallel$$\mathbf{c}$, thus, the strong anisotropy forces the spins to lie within the $ab$ plane.

In CaBaCo$_4$O$_7$, symmetry allows TD when the light propagates along the toroidal moment, {\bf T}={\bf P}$\times${\bf M}, where the electric polarization {\bf P} points along the $c$ axis and the magnetization {\bf M} is within the easy-plane. Since the symmetry reduces from $Pbn2_1$ to $Pb'n2'_1$ upon the paramagnetic to ferrimagnetic phase transition, two magnetic domains interchanged by the time reversal operation can exist, Fig.~\ref{fig1}. The domains possess the same electric polarization, while their magnetization and thus their toroidal moments have opposite signs. The finite magnetization of the domains allows to switch between them by an external magnetic field larger than about 2\,T at $T$=5\,K. Due to the strong exchange coupling between the spins and the small in-plane anisotropy, we expect that the toroidal moment formed by the polar $c$ axis and a magnetization pointing to any in-plane directions can be reversed by reversing the external magnetic field. Therefore, instead of reversing the position of the source and the detector, which would be the time reversed experiment, we measured the absorption difference in positive and negative in-plane magnetic fields while keeping the optical system intact in order to sensitively detect the directional dichroism. In this way, we reversed the toroidal moment of the sample instead of the light propagation vector, as TD depends only on their relative orientation.

\begin{figure}[t!]
\includegraphics[width=3in]{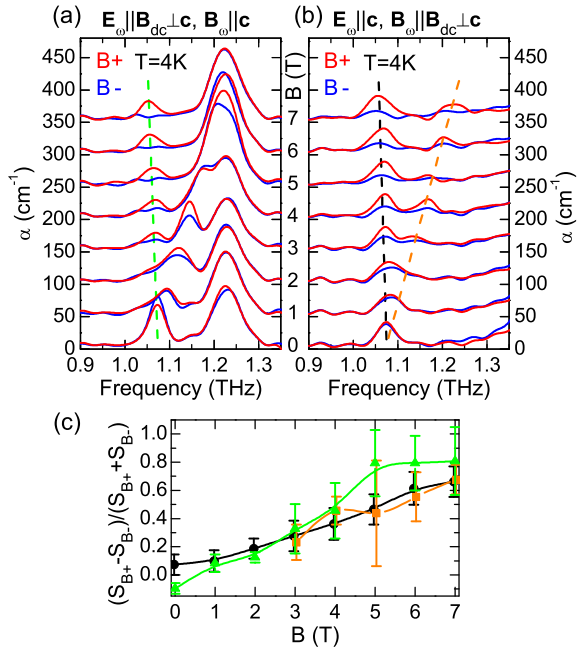}\\
\caption{(Color online) (a)-(b) The magnetic field dependence of the THz absorption spectra in CaBaCo$_4$O$_7$ measured for light propagation along the toroidal moment, {\bf T}={\bf P}$\times${\bf M} formed by the cross product of the magnetization, {\bf M} and the electric polarization, {\bf P}. Upon the reversal of the magnetic field the toroidal moment is reversed which results in strong non-reciprocal directional dichroism. (c) The magnetic field dependence of the dissymmetry factor (defined in the text). The colors of the symbols corresponds to those of the dashed lines in panels (a) and (b) showing the field evolution of the resonance frequencies. In panel (c) the solid lines are guides to the eyes.}
\label{fig4}
\end{figure}

The magnetic field dependence of the absorption spectra both for positive and negative in-plane fields, was measured using time-domain THz spectroscopy in Voigt geometry, {\bf k}$\perp${\bf B}$_{dc}$, for light propagation along the toroidal moment {\bf T} (see Fig.~\ref{fig4}). In this setup the highest field was limited to $\pm$7\,T, and the high frequency cut-off was below 1.4\,THz. Fig.~\ref{fig4} (a) and (b) present two sets of spectra measured with the two orthogonal light polarizations, {\bf E}$_\omega$$\perp$$\mathbf{c}$ and {\bf E}$_\omega$$\parallel$$\mathbf{c}$, respectively. In order to study a single magnetic domain, first the highest positive (negative) magnetic field, +7\,T (-7\,T) was applied to reach the saturation magnetization, and the magnetic field dependence was measured with decreasing the field down to zero. Remarkably, the lower frequency branch of the 1.07\,THz magnetoelectric mode can be excited only in positive fields above +5\,T for light polarization {\bf E}$_\omega$$\perp$$\mathbf{c}$. Since this mode absorbs light only for one of the propagation directions with respect to the toroidal moment and it is transparent for the other, CaBaCo$_4$O$_7$ allows unidirectional light propagation around the resonance frequency of the magnetoelectric mode.

The strength of the directional dichroism is quantified by the dissymmetry factor defined as $\eta$=($S$$_{B+}$-$S$$_{B-}$)/($S$$_{B+}$+$S$$_{B-}$), where $S$$_{B+}$ and $S$$_{B-}$ are the strength of the resonance in positive and negative fields, respectively \cite{Barron2004}. First, $S$$_{B+}$ was determined by fitting the absorption spectrum with a sum of Lorentzian functions, then the negative field spectra were fitted by allowing only the strength parameter, $S$$_{B-}$ to be varied. For the two light polarizations the dissymmetry factor has different magnetic field dependencies as shown in Fig.\ref{fig4} (c). When {\bf E}$_\omega$$\perp$$\mathbf{c}$ the TD is saturated close to the theoretical maximum $\eta$=1 above 5\,T for the lower branch of the 1.07\,THz magnetoelectric mode. However, for the orthogonal polarization {\bf E}$_\omega$$\parallel$$\mathbf{c}$, the dissymmetry factor monotonously increases in the studied field range. The difference in the field dependencies is likely caused by the different hybridizations between the magnon mode and higher energy polar modes governed by different selection rules for the two light polarizations.

As it was shown previously~\cite{Kezsmarki2014}, such a strong directional dichroism, $\eta$$\approx$1, can be achieved for a magnetoelectric resonance when the ratio of the magnetic and electric dipole matrix elements, $\langle$n$|$m$|$0$\rangle$ and $\langle$n$|$p$|$0$\rangle$, respectively, is equal to the speed of light within the material for the transparent direction:
\begin{equation}
\frac{\langle n|m|0\rangle}{\langle n|p|0\rangle}=\frac{c_0}{\sqrt{\varepsilon}}.
\label{EqUniDir}
\end{equation}
Here $c$$_0$ is the speed of light in vacuum and $\varepsilon$ is the background dielectric constant. We get $\varepsilon$=14.7 from our time-domain measurement, thus, the ratio, Eq.\ref{EqUniDir}, is $c$$_0$/3.7 for the 1.07\,THz mode of the CaBaCo$_4$O$_7$ above 5\,T and {\bf E}$_\omega$$\perp$$\mathbf{c}$, {\bf B}$_\omega$$\parallel$$\mathbf{c}$.

From a classical spin-wave theory one expects maximum 16 magnon modes in CaBaCo$_4$O$_7$ as the unit cell contains 16 magnetically different atoms. In the Faraday geometry (see Fig.\ref{fig3}) three modes are observed with resonance frequencies monotonously increasing in field. In the Voigt geometry (see Fig.\ref{fig4}) the 1.07\,THz mode is split into a magnetic field independent mode and a higher frequency branch, which is also observed in the Faraday configuration. This raises the number of observed magnon modes to four. The missing spin excitations are outside the frequency window of the current Faraday and Voigt experiments or they do not interact with the electromagnetic radiation.

%\section{Conclusions}

In conclusion, we have studied the magnetic resonances in the type-I multiferroic CaBaCo$_4$O$_7$. The lowest-energy mode at 1.07\,THz is a mixed magnetic and electric dipole resonance, while the 1.41\,THz resonance can be excited only by the electric field component of the light. Such magnetoelectric and electric dipole modes are probably allowed by the strong inversion symmetry breaking fields of the present pyroelectric compound. For counter-propagating light beams, the 1.07\,THz resonance absorbs the electromagnetic radiation only for one propagation direction due to the dynamic magnetoelectric coupling. Nearly perfect unidirectional light absorption is realized in CaBaCo$_4$O$_7$, and the transparent/absorbing directions can be controlled by an external magnetic field; our findings can pave a way for future applications of mutliferroics in diodes for terahertz light. Furthermore, the magnetic field dependence of the spin excitations was determined in a broad magnetic field and energy region, which provides valuable information for the construction of a microscopic spin model of CaBaCo$_4$O$_7$, and can help to understand the large magnetic order induced polarization.

%\section*{Acknowledgement}

We are grateful to I. K\'ezsm\'arki for fruitful discussions. This work was supported by the Funding Program for World-Leading Innovative R\&D on Science and Technology (FIRST Program), Japan, by Hungarian Research Funds OTKA K 108918, OTKA PD 111756, Bolyai 00565/14/11 and by the Lend\"ulet Program of the Hungarian Academy of Sciences, and by the Estonian Ministry of Education and Research Grant IUT23-03 and by the Estonian Science Foundation Grant ETF8703.


\begin{thebibliography}{99}
\bibitem{Gajek2007} M. Gajek, M. Bibes, S. Fusil, K. Bouzehouane, J. Fontcuberta, A. Barth\'el\'emy, and A. Fert, Nature Materials {\bf 6}, 296 (2007).
\bibitem{Matsukura2015} F. Matsukura, Y. Tokura, and H. Ohno, Nature Nanotechnology {\bf 10}, 209 (2015).
\bibitem{Wang2009}K.F. Wang, J.-M. Liu, and Z.F. Ren, Advances in Physics {\bf 58}, 321 (2009).
\bibitem{Jung2004} J.H. Jung, {\it et al.}, \prl {\bf 93}, 037403 (2004).
\bibitem{Kubota2004} M. Kubota, {\it et al.}, \prl {\bf 92}, 137401 (2004).
\bibitem{Arima2008Ome} M. Saito, K. Taniguchi, and T. Arima, J. Phys. Soc. Jpn. {\bf 77}, 013705 (2008).
\bibitem{Kezsmarki2011}
I. K\'ezsm\'arki, N. Kida, H. Murakawa, S. Bord\'acs, Y. Onose, and Y. Tokura, Phys. Rev. Lett. {\bf 106}, 057403 (2011).
\bibitem{Bordacs2012}
S. Bord\'acs, {\it et al.}, Nat. Phys. {\bf 8}, 734 (2012).
\bibitem{Takahashi2012}
Y. Takahashi, R. Shimano, Y. Kaneko, H. Murakawa, and Y. Tokura, Nat. Phys. {\bf 8}, 121 (2012).
\bibitem{Takahashi2013}
Y. Takahashi, Y. Yamasaki, and Y. Tokura, Phys. Rev. Lett. {\bf 111}, 037204 (2013).
\bibitem{Kezsmarki2014}
I. K\'ezsm\'arki, {\it et al.}, Nat. Commun. {\bf 5}, 3203 (2014).
\bibitem{Kibayashi2014}
S. Kibayashi, Y. Takahashi, S. Seki, and Y. Tokura, Nat. Commun. {\bf 5}, 4583 (2014).
\bibitem{Kuzmenko2014} A. M. Kuzmenko, {\it et al.}, Phys. Rev. B {\bf 89}, 174407 (2014).
\bibitem{KezsmarkiBFO2015}I. K\'ezsm\'arki, U. Nagel, S. Bord\'acs, R.S. Fishman, J.H. Lee, H.T. Yi, S.W. Cheong, and T. R\~o\~om, Phys. Rev. Lett. {\bf 115}, 127203 (2015).
\bibitem{Tamagnone2014} M. Tamagnone, A. Fallahi, J. R. Mosig, and J. Perruisseau-Carrier, Nature Photon. {\bf 8}, 556 (2014).
\bibitem{Szaller2014}D. Szaller, S. Bord\'acs, V. Kocsis, T. R\~o\~om, U. Nagel, and I. K\'ezsm\'arki, Phys. Rev. B {\bf 89}, 184419 (2014).
\bibitem{Caignaert2013} V. Caignaert, A. Maignan, K. Singh, Ch. Simon, V. Pralong, B. Raveau, J. F. Mitchell, H. Zheng, A. Huq, and L. C. Chapon, Phys. Rev. B {\bf 88}, 174403 (2013).
\bibitem{Barron2004} L. D. Barron, {\it Molecular light scattering and optical activity}, (Cambridge University Press, 2004).
\bibitem{note1} To be precise, when the ferroic order of the local toroidal moments are present, the TD can be observed even if the macroscopic magnetizzation and/or polarization vanish.
\bibitem{Raveau2008} B. Raveau, V. Caignaert, V. Pralong, D. Pelloquin, and A. Maignan, Chem. Mater {\bf 20}, 6295 (2008).
\bibitem{Caignaert2009}
V. Caignaert, V. Pralong, A. Maignan, and B. Raveau, Solid State Commun. {\bf 149}, 453 (2009).
\bibitem{Caignaert2010}
V. Caignaert, V. Pralong, V. Hardy, C. Ritter, and B. Raveau, Phys. Rev. B {\bf 81}, 094417 (2010).
\bibitem{Iwamoto2012}
H. Iwamoto, M. Ehara, M. Akaki, and H. Kuwahara, J. Phys.: Conf. Ser. {\bf 400}, 032031 (2012).
\bibitem{Johnson2014}
R. D. Johnson, K. Cao, F. Giustino, and P. G. Radaelli, Phys. Rev. B {\bf 90}, 045129 (2014).
\bibitem{Kida2004} N. Kida, {\it et al.}, Phys. Rev. B {\bf 78}, 104414 (2008).
\bibitem{Room2004}
T. R\~o\~om, D. H\"uvonen, U. Nagel, Y.-J. Wang, and R. K. Kremer, Phys. Rev. B {\bf 69}, 144410 (2004).
\bibitem{Pimenov2006} A. Pimenov, A. A. Mukhin, V. Yu. Ivanov, V. D. Travkin, A. M. Balbashov, and A. Loidl, Nat. Phys. {\bf 2}, 97 (2006).
\end{thebibliography}
\end{document}